\documentclass[doublecol,linenumbers]{epl2}

\title{Effective spin-fermion model for strongly correlated electrons}

\author{A. Ferraz\inst{1} \and E. Kochetov\inst{1,2}}

\institute{
  \inst{1} International Institute of Physics - UFRN,
Department of Experimental and Theoretical Physics - UFRN, Natal, Brazil \\
  \inst{2} Laboratory of Theoretical Physics, Joint
Institute for Nuclear Research, 141980 Dubna, Russia}


\abstract{A  modified spin-fermion model is proposed to describe
the physics of the underdoped phase of the $t-J$ model of strongly correlated electrons.}

\begin{document}

\maketitle

More than a decade ago, a spin-fermion (SF) model was put forward in an attempt to account for the observed non-Fermi liquid anomalies in the doped cuprates near optimal doping.\cite{b.chub} It describes low-energy fermions with a large Fermi surface (FS) interacting
with each other via soft collective spin excitations. The bare spin propagator emerges from the integration of the high-energy fermionic modes,
\begin{equation}
\chi_0(\vec q,\omega_n)\sim (\frac{\omega_n^2}{v_s^{2}} +(\vec q-\vec Q)^2+ \xi^{-2}_{AF})^{-1},
\label{bsp}\end{equation}
where $\omega_n=2\pi n/\beta$ are  Matsubara bosonic frequencies.
The antiferromagnetic (AF) spin correlation length, $\xi_{AF}$,  and the spin velocity, $v_s$,
are the high-energy phenomenological parameters.
The spin velocity is of order of the Fermi velocity, $v_F.$

At criticality $(\xi_{AF} =\infty)$, the singular nature of the low-energy AF paramagnons singles out the so-called {\it hotspots} on the FS. These are the points on the FS for which the AF momentum $\vec Q=\vec Q_{AF}=(\pi,
\pi)$ coincides with $2\vec k_F$.
Due to the enormous enhancement of the AF fluctuations at or near the quantum critical point (QCP),
the fermions in the vicinity of hotspots interact strongly with each other via singular AF paramagnons. This in turn results in a strong damping of the propagating spin modes in the low-energy limit. The spin modes become dissipative and the fermion quasiparticles acquire a short lifetime. As a result, the electron system displays
a singular non-FL behavior at low energies.\cite{b.chub, b.alt}

There are, however, two necessary conditions for this approach to work.
First, a SF model must be at weak coupling. As a result, a spin-fermion coupling $g$ should be smaller than the
fermionic bandwidth. Only in this case can one perturbatively integrate out high energies to obtain an  effective low-energy action in which the spin degrees of freedom are represented by the bare spin propagator (\ref{bsp}). Second, the low-energy one-loop spin-fermion dynamics renormalizes the bare spin propagator producing the effective susceptibility,
\begin{equation}
\chi_{eff}(\vec q,\omega_n)\sim (\frac{\omega_n^2}{v_s^{2}} +(\vec q-\vec Q)^2+ \xi^{-2}_{AF}+\gamma|\omega_n|)^{-1},
\label{esp}\end{equation}
where a dissipation coefficient $\gamma\sim g^2.$
The non-FL low-energy behaviour sets in provided $\gamma$ is large enough
to allow for the  dropping of the "ballistic" piece $\frac{\omega_n^2}{v_s^{2}}$ in the presence of a much more pronounced "dissipative" contribution.
These two conditions formally contradict each other, however.

A possible way out of that is to  employ a $1/N$ expansion, where $N$ stands for a number of the fermion flavours with $N=1$ being the physical case.  It can be shown that a random phase approximation (RPA) generated by a one-loop particle-hole expansion contributes to the order $O(1)$ whereas all other higher loop diagrams vanish in the limit $N\to\infty$. At the same time, $\gamma$ remains large and finite. The $1/N$ is then the parameter that keeps the theory under control.
The SF theory was developed in detail, along these lines, in a seemingly self-consistent way in Ref.\cite{b.chub}. A good agreement with experiment was declared as well.

However, such a theory is incomplete.
The $1/N$ approach can only be applied reliably, provided the physics remains qualitatively the same at any $N$.
This is not the case in the present situation. Moreover, the $1/N$ approximation fails to take proper account of the non-FL
physics displayed by the underdoped cuprates.\cite{b.tsvelik}
More specifically, it was recently shown that the $1/N$ approach to a $2+1$ SF model misses some important contribution in both perturbation and  renormalization group theory \cite{b.lee, b.metl}, which invalidates its application in this context. The proposed $1/N$ SF model is in this way not well defined and this makes a direct comparison with experiment incomplete and unjustified.

A new interesting approach to treat a strong Landau damping within a framework of a weakly coupled SF model in a controlled way was recently proposed in \cite{b.efetov}. Instead of the introduction of the $N$ fermion flavours, the authors consider a reduced FS to account for the low-energy dynamics of the optimally doped cuprates. Such a FS comprises eight hotspots which are essentially
the intersects of the non-interacting hole-like FS with the magnetic zone boundary.\cite{b.metl} This zone boundary is fully sensitive to umklapp scattering processes which gap out the FS completely at half-filling. A small angle $\delta$  between the Fermi velocities of two hot spots involved in scattering processes is introduced in this new approach. The Landau damping modifies the form of the bare spin propagator as given by (\ref{esp}) with $\gamma$ being now given by
$\gamma \sim g^2/sin\delta.$ In the limit $\delta\ll 1$, the Landau damping is strong even within a "weakly coupled" SF theory. This keeps the theory under control and it is, in principle, in agreement with experimental data. In particular, the proposed theory features
an instability towards a formation of a quadropole charge order (CO).
The magnetic-mediated interaction is shown to give rise
to charge bond order with the momentum directed along the Brillouin zone diagonal. While the structure factor is compatible with experiments, the magnitude and direction of the ordering momentum are not.
In experiments, the charge ordering momentum appears to be always directed along either the horizontal or the vertical axis.

By construction, the approach advocated in \cite{b.efetov} relies on a specific shape of a reduced hotspot FS
which should describe the cuprates near optimal doping. Apart from the already mentioned stability problems of such a FS,
it is not clear if this approach is applicable at lower dopings in which strong electron-electron interactions are known to be at work. The heavily underdoped region is characterized by strong electron correlations which keep the majority of the electrons well localized.
Accordingly, the  FS consists of small separated patches (electron pockets).
It is still unclear how far a standard weak-coupling SF model can go to describe realistically
the physics in this strongly correlated region as well.

The goal of the present paper is to explore this issue further. We show that, at least close to half-filling,
the standard SF model should be drastically modified to take into account strong electron correlations -- a hallmark
of the physics of the underdoped cuprates.

The SF model can be derived formally \cite{b.3} from the Hubbard model which describes fermions hopping on a lattice with a tunneling amplitude $t_{ij}$ and are subjected to a short-range Coulomb repulsive interaction $U$:
\begin{equation}
H_{tU} = \sum_{\vec k\sigma} t_{\vec k} c_{\vec k\sigma}^{\dagger}
c_{\vec k\sigma} + U\sum_i n_{i\uparrow}n_{i\downarrow}.
\label{hubb}\end{equation}
Operator $c_{i\sigma}^{\dagger}$ is a creation operator for electrons with spins $\sigma$ on site $i$
and $n_{i\sigma}=c_{i\sigma}^{\dagger}c_{i\sigma}$ is the on-site electron number operator
with the spin projection $\sigma$.
The next basic step is the Hubbard-Stratonovich (HS) decoupling of the quartic fermion interaction.
The SF model is then described by the Lagrangian \cite{b.3}
\begin{eqnarray}
L_{tg}^{SF}&=&
\sum_{ q\sigma}\bar c_{q\sigma}(-i\omega_n+t_{\vec q})c_{q\sigma}
\nonumber\\
&+&
\sum_{q}\chi^{-1}_0(q) \vec \phi_{\vec q}\vec \phi_{-\vec q}+g\sum_{q}
\vec s_{\vec q-\vec Q}\vec \phi_{-\vec q}.
\label{sf}\end{eqnarray}
where  $g\sim U$ is a spin-fermion coupling constant.
The electron spin operator
\begin{equation}
\vec s_{i}=\sum_{\sigma,\sigma'}c^{\dagger}_{i\sigma}\vec \tau_{\sigma\sigma'} c_{i\sigma'},
\label{q}\end{equation}
couples to the auxiliary HS field $\vec \phi_{i}.$
In case the SF coupling $g\sim U$ is smaller than the fermionic bandwidth, $W\sim t$, one can apply a standard perturbation theory controlled by a small parameter $U/t$.

In the cuprates, however, the fermionic bandwidth is comparable to the
Coulomb repulsion and this makes the weak-coupling theory problematic.
Besides, close to half filling,
the tunnelling amplitude gets effectively renormalized, i.e.,
$t\to t_{eff}\sim xt$, where $x \ll 1$ is the doping concentration.\cite{b.bza}
This results in an extremely narrow bandwidth for the itinerant electrons
which automatically become subject to the important (in this region) local no double electron occupancy (NDO) constraint.
This makes the hopping of an electron on an already occupied
lattice site extremely unfavorable since $U\gg xt.$
This results in extra strong correlations between electrons in addition to those caused by the Pauli exclusion principle.
The lightly doped cuprates are thus characterized by the strong electron correlations in the coupling regime $U/t_{eff}\gg 1$.
Since the SF coupling $g\sim U$, a straightforward extension of Eq.(\ref{sf})
in this regime automatically implies that $g$ becomes large.

Let us now see what happens in the limit of an infinitely strong Coulomb coupling.
The Hubbard model goes over into
\begin{equation}
H_{U=\infty}=\sum_{ij\sigma} t_{ij} \tilde c_{i\sigma}^{\dagger}
\tilde c_{j\sigma},
\label{inf}\end{equation}
where the $\tilde c_{i\sigma}$s are the on-site Gutzwiller projected electorn operators:
\begin{equation}
\tilde c_{i\sigma}=P_ic_{i\sigma}P_i, \,\,\,\, P_i=1-n_{i\uparrow}n_{i\downarrow}.
\label{pel}\end{equation}
The model (\ref{inf}) exhibits the extreme case of strong correlations and  is referred to as the
$U=\infty$ Hubbard model.
This model is certainly far from
trivial and it admits an exact solution only in $1d$.
Under the assumption that, in the limit $g\to\infty$, the SF model (\ref{sf}) is still meaningful, it must recover in this limit the physics displayed by Eq.(\ref{inf}). However, the limit $g\to +\infty$ pushes the energy levels of model (\ref{sf})
to $+\infty$ and this destabilizes the system. We should therefore search for a modified SF
model which is finite in the limit of strong electron correlations.

To deal with the underdoped phase, it is more convenient to use as a reference framework the so-called $t-J$ model of strongly
correlated electrons:
\begin{eqnarray}
H_{tJ}=\sum_{ij\sigma} t_{ij} \tilde c_{i\sigma}^{\dagger}
\tilde c_{j\sigma} + J\sum_{ij}\vec s_i\vec sj,
\label{tJ}\end{eqnarray}
where $J=4t^2/U$.
The $t-J$ model is essentially the strong coupling limit of the Hubbard model, i.e.,
$H_{tU}=H_{tJ} +  o(J/t),\,t/U\ll 1.$ It is believed to capture the basic physics of strongly correlated electrons
in the underdoped cuprates.

Within the recently proposed spin-dopon representation of the $t-J$ model \cite{b.pfk},
the local NDO constraint is enforced
by the requirement of an infinitely large spin-dopon coupling.
In this way, the $t-J$ model turns out to be equivalent to a Kondo-Heisenberg lattice model
of itinerant dopons (doped holes) and
localized lattice spins (localized electrons) at an infinite Kondo coupling:
\begin{equation} H_{tJ} = \sum_{ij\sigma}
T_{ij}{d}_{i\sigma}^{\dagger} {d}_{j\sigma}+
J\sum_{ij} \vec S_i \cdot \vec S_j +\lambda
\sum_{i}\vec{S_i} \cdot \vec s_i,
\label{ssf}\end{equation}
where $T_{ij}=2t_{ij}+(3\lambda/4-\mu)\delta_{ij}$ and the
$d_{i\sigma}$'s are the fermion operators to describe the dopons. The $su(2)$ generators $\vec S_i$ describe the localized electron spins. The on-site itinerant dopon spin operator is given by Eq.(\ref{q}) where the electron operators being replaced by the dopon operators.
A global parameter $\lambda$ is to be sent to $+\infty$ to ensure the selection of the appropriate physical subspace.
The unphysical doubly occupied electron states are separated from the physical
sector by an energy gap $\sim\lambda$.
In the $\lambda\to +\infty$ limit, i.e. in the limit that $\lambda$ is much larger than any other existing scale in the problem, those states
are  automatically excluded from the Hilbert space.

In this limit, the high and low energy itinerant fermions cannot be separated out and this is another manifestation of the Mott physics.
One cannot therefore integrate out high-energy fermions in that case. We assume however that this separation still holds
in the spin-dopon representation for well localized
particles, i.e., the localized spin degrees of freedom.
\revision{After all, this is believed to be the case for 
the undoped Mott insulator (the quantum Heisenberg model) - a $2d$ lattice of  localized electrons. 
We believe that such a separation is valid at sufficiently low doping as well, since a large majority of the electrons remain
localized in this regime.}

At half-filling, $x=0$, and model (\ref{ssf}) reduces to the $2d$ AF Heisenberg model,
$$J\sum_{ij} \vec S_i \cdot \vec S_j, \quad J> 0.$$
This model exhibits long-range AF order at $T=0$. However small dopings $(x_c\approx 0.05)$ suffice to destroy that order leaving behind strong short-range AF fluctuations that persist up to optimal doping. This can be taken into account as follows.
We write
\begin{equation}
\vec S_i(\tau)=(-)^i\vec n(x_i,\tau)+\vec m(x_i,\tau),
\label{anz}\end{equation}
where $\vec n(x_i,\tau)$ is a smooth function of $(x,t)$ with $|\vec n(x_i,\tau)|=1$. Here $\vec n$ describes the low energy AF fluctuations. The vector field $\vec m$ with $|\vec m(x_i,\tau)|\ll 1$ and $\vec n(x_i,\tau)\cdot\vec m(x_i,\tau)=0$ describes the ferromagnetic fluctuations which should in turn be integrated out. This approach is controlled by a $1/2s$ expansion, where $s$ is the electron spin.
Although the physical case corresponds to $s=1/2,$ the physics of the short-range $2d$ AF phase is correctly reproduced in this case.

We then substitute the ansatz (\ref{anz}) into (\ref{ssf}) and drop the small $\vec m$-dependent term in the spin-dopon interaction,
\begin{equation}
\lambda\sum_{i}\vec S_i\cdot\vec s_i\to
\lambda\sum_{\vec q} \vec n_{\vec q-\vec Q} \cdot \vec s_{-\vec q}.
\label{inp}\end{equation}
The high-energy integration over $\vec m$ can then be readily done in a standard fashion \cite{b.haldane} resulting in a SF Lagrangian to describe strongly correlated electrons in the underdoped phase,
\begin{eqnarray}
L_{t\lambda}^{SF}&=&
\sum_{ q\sigma}\bar d_{q\sigma}(-i\omega_n+T_{\vec q}(\lambda))d_{q\sigma}
\nonumber\\
&+&
\sum_{q}D^{-1}_0(q)\vec n_{\vec q}\vec n_{-\vec q}+\lambda\sum_{q}
\vec s_{\vec q-\vec Q}\vec n_{-\vec q}.
\label{f}\end{eqnarray}
Here $q=(\vec q, i\omega_n)$ and
\begin{equation}
T_{\vec q}(\lambda)=2t_{\vec q}+\frac{3\lambda}{4}-\mu.
\label{T}\end{equation}
The spin propagator $D(q)$ is given by Eq.(\ref{bsp}), where
$v_s\sim Ja$ is now the spin-wave velocity, and $a$ is a lattice spacing.

We thus arrive at the SF model at an infinitely strong coupling $\lambda$. In contrast with the standard model (\ref{sf}), 
the renormalized hopping amplitude contains the spin-fermion coupling $\lambda$ as well, as given by  Eq.(\ref{T}).
Because of this, in the physical limit, $\lambda\to +\infty$, the model (\ref{f}) remains finite.
In $1d$ it reproduces the exact ground-state energy of Hamiltonian (\ref{inf}).\cite{b.fk}

It is instructive to consider (\ref{f}) at small values of $\lambda, \, \lambda/t\ll 1.$
This is  clearly a unphysical assumption in the underdoped region and in the whole PG phase as well. 
 In this hypothetical case, one can ignore the $\lambda$-dependence of the hopping amplitude $T_{ij}$. 
Equation (\ref{f}) then reduces to the weak-coupling SF Lagrangian  given by Eq.(\ref{sf})
with the identifications $g\to \lambda$ and $t_{\vec k} \to 2t_{\vec k}.$
We thus see that ignoring strong electron correlations indeed results in the standard weak-coupling SF model.
There is however a qualitative distinction between the standard representation (\ref{sf}) and that given by 
our equation (\ref{f}).
The spin velocity in Eq.(\ref{bsp}) is of order of the Fermi velocity $v_F\sim ta$, whereas the spin-wave velocity $v_s$ in
$D(q)$ is determined by the dynamics of the localized spins, i.e., $v_s\sim Ja$.

\revision{As a result, the correct energy scale in the problem is the spin fluctuation frequency, $\omega_{sf}\sim v_s  \xi^{-1}/\kappa,$ 
where $\kappa$ is an effective dimensionless 
spin-fermiom coupling which is essentially
$\sim \lambda\xi/v_s,$ and $\xi:= \xi_{AF}$. This frequency measures  
the deviation from the QCP.  At the QCP ($\xi=\infty$), 
$\omega_{sf}=0$, and 
the system displays non-Fermi-liquid behavior down to the quasiparticle fermionic energy $\omega=0$.
Away from the QCP, the correlation length $\xi$ is finite. For small energies, $\omega << \omega_{sf}$, 
the system displays  FL behavior, although with a damping term scaling inversely with $\omega_{sf}$, rather than with the Fermi
energy as in conventional metals. At frequencies above $\omega_{sf}$,  the system crosses over into the 
non-Fermi liquid quantum-critical regime.
In the conventional SF approach,  $\omega_{sf}\sim v_F^2/\xi^2,$
whereas in our case  $\omega_{sf}\sim v_s^2/\xi^2.$
Since $\omega_{sf}(v_s)/\omega_{sf}(v_F)=(J/t)^2<<1$, this implies 
that the energy window within which the essential non-Fermi-liquid behavior sets in
is in fact much larger than that predicted by the conventional SF approach, solely based on the 
itinerant fermion description.}

To conclude, we propose an effective SF model that takes into account strong electron correlations.
These correlations bring us naturally to the SF model at infinitely strong coupling with a simultaneously renormalized hopping
amplitude. This theory is difficult to implement analytically, since there is still no machinery available with the proper technical tools.
However, it is just this effective model that is appropriate to address the cuprates at very low doping, rather than its weak-coupling SF version. The latter leaves out the effects produced by the NDO -- the essence of the physics of
strongly correlated electrons.

\acknowledgments

This work was partly supported by the MCTI, MEC and CNPq (Brasil).

\end{document}